\documentclass{article}
\usepackage{amssymb}
\usepackage{amsfonts}
\usepackage{graphicx}
\usepackage{amsmath}
\usepackage{geometry}
\usepackage{scalefnt}
\usepackage{ulem}
\usepackage{hyperref}
\usepackage{slashed}
\usepackage{multicol}
\usepackage{indentfirst}

\input{tcilatex}

\begin{document}

\title{Deformation Quantization, Quantization, and the Klein-Gordon Equation}
\author{Philip Tillman \\
%EndAName
Department of Physics and Astronomy, University of Pittsburgh, Pittsburgh,
PA, USA\\
phil.tillman@gmail.com}
\date{\today }
\maketitle

\begin{abstract}
The aim of this proceeding is to give a basic introduction to Deformation
Quantization (DQ) to physicists. We compare DQ to canonical quantization and
path integral methods. It is described how certain issues such as the roles
of associativity, covariance, dynamics, and operator orderings are
understood in the context of DQ. Convergence issues in DQ are mentioned.
Additionally, we formulate the Klein-Gordon (KG) equation in DQ. Original
results are discussed which include the exact construction of the Fedosov
star-product on the dS and AdS space-times. Also, the KG equation is written
down for these space-times.

This is a proceedings to the Second International Conference on Quantum
Theories and Renormalization Group in Gravity and Cosmology.
\end{abstract}

%TCIMACRO{\TeXButton{\begin{multicols}{2}}{\begin{multicols}{2}}}%
%BeginExpansion
\begin{multicols}{2}%
%EndExpansion

\section{Introduction}

In this report we discuss several issues regarding quantization and how some
of them can be better understood by using deformation quantization (DQ).
These issues include the role of covariance and associativity in canonical
quantization, and the role of the Lagrangian in the path integral method.
Another issue discussed is the operator ordering ambiguities in quantization
in the context of DQ

In addition, we illustrate how to write down the Klein-Gordon (KG) equation
in DQ, and how to move back and forth from Hilbert space representations to
DQ. It is verified, for the case of dS and AdS, that this KG equation and
algebra of observables yield the standard results (see, for example, 
%TCIMACRO{%
%\TeXButton{Fr\o nsdal C. 1965, 1973, 1975a, 1975b}{\hyperlink{ref1}{Fr\o nsdal C. 1965, 1973, 1975a, 1975b}}}%
%BeginExpansion
\hyperlink{ref1}{Fr\o nsdal C. 1965, 1973, 1975a, 1975b}%
%EndExpansion
).

The main problem in DQ, as I see it, is related to the standard treatments
of deformation products which rely heavily on series expansions in a formal
parameter $\hbar $. To partially address the convergence of these series,
the Fedosov star-product (a generalization of the Groenewold-Moyal
star-product) is computed exactly for the examples of the dS and AdS
space-times. Once the star-algebra is computed, the Klein-Gordon (KG)
equation is then calculated.

\section{Quantization on Space-Times}

This section is a brief summary of some important issues (which can be
confusing) about how to properly construct quantum theories on space-times
using canonical quantization, path integral methods, and DQ. This is an
attempt to ascertain some of the essential features of quantum theories. We
begin with canonical quantization formulated by Dirac.

\subsection{Canonical Quantization}

The Dirac canonical quantization map $Q$ is a map that tries to assign to
each phase-space function $f$ an operator $Q\left( f\right) $ (also denoted
by $\hat{f}$) that acts on an appropriate Hilbert space. $Q$ is defined by
the following four properties:

\begin{center}
\begin{tabular}{ll}
1. & $Q\left( c_{1}f+c_{2}g\right) =c_{1}Q\left( f\right) +c_{2}Q\left(
g\right) $ \\ 
2. & $Q\left( \left\{ f,g\right\} \right) =\left[ Q\left( f\right) ,Q\left(
g\right) \right] /i\hbar $ \\ 
3. & $Q\left( 1\right) =I$ \\ 
4. & $Q\left( x\right) ,Q\left( p\right) \text{ are represented irreducibly}$%
\end{tabular}
\end{center}

\noindent for all constants $c_{1},c_{2}\in 
%TCIMACRO{\U{2102} }%
%BeginExpansion
\mathbb{C}
%EndExpansion
$, $\left\{ ,\right\} $ is the Poisson bracket, and where $I$ is the unit
element in the algebra.

However there is a major problem with the above setup. The theorem of
Groenewold and van Howe states that a consistent quantum theory satisfying
rules $1$ through $4$ is impossible.

It can be seen easily seen that property 2 is inconsistent by trying to
quantize the function $9x^{2}p^{2}$ in two ways. One using $%
9x^{2}p^{2}=\left\{ x^{3},p^{3}\right\} $ and the other using $9x^{2}p^{2}=\{%
\sqrt{3}x^{2}p,\sqrt{3}xp^{2}\}$. You will see that you obtain two different
values for $Q\left( 9x^{2}p^{2}\right) $ which is a contradiction.

We can get around this "no go" thereom by running the procedure for
functions that are at most quadratic in the phase-space variables $x$ and $p$
(see 
%TCIMACRO{\TeXButton{Giulini D. 2003}{\hyperlink{ref1}{Giulini D. 2003}}}%
%BeginExpansion
\hyperlink{ref1}{Giulini D. 2003}%
%EndExpansion
). The resulting elements $\left\{ Q\left( x\right) ,Q\left( p\right)
,Q\left( x^{2}\right) ,Q\left( p^{2}\right) ,Q\left( xp\right) \right\} $
are forced to form the basis of an associative operator algebra which
becomes our observable algebra. The procedure $Q$, subsequently, is
consistent \textit{only} on this subset. Therefore, standard canonical
quantization is understood in these terms, by the quantization of these
basic variables $\left( x,p\right) $. The main problem is that the procedure
seems to depend on which coordinates $\left( x,p\right) $ you choose.

There are ways to get around this problem by modifying the properties above.
DQ solves the inconsistency by modifying property 2. This is achieved by
forcing associativity of the resulting algebra of observables (see 
%TCIMACRO{%
%\TeXButton{Gozzi E. and Reuter M. 1994}{\hyperlink{ref1}{Gozzi E. and Reuter M. 1994}}}%
%BeginExpansion
\hyperlink{ref1}{Gozzi E. and Reuter M. 1994}%
%EndExpansion
). Another way of fixing this problem is by abandoning property 4 which is
the approach adopted by prequantization in geometric quantization. To go
from prequantization to full quantization in general is an unsolved problem
in geometric quantization (see 
%TCIMACRO{%
%\TeXButton{Woodhouse N. 1980}{\hyperlink{ref1}{Woodhouse N. 1980}}}%
%BeginExpansion
\hyperlink{ref1}{Woodhouse N. 1980}%
%EndExpansion
).

The main reason for abandoning property 2 is that it is inconsistent with
associativity. First we start with a definition:

\textbf{Def. }A \textbf{Poisson algebra} is any algebra equipped on
phase-space with a product $C\left( \cdot ,\cdot \right) $ where the
antisymmetric part of $C$ for any functions $f$ and $g$ is the Poisson
bracket:%
\begin{equation*}
C\left( f,g\right) -C\left( g,f\right) =\left[ f,g\right] _{P}
\end{equation*}%
The identity element is $1$:%
\begin{equation*}
C\left( 1,f\right) =C\left( f,1\right) =f
\end{equation*}%
An example of a Poisson algebra is $C\left( f,g\right) =1+\frac{1}{2}\left[
f,g\right] _{P}$.

A Poisson algebra is necessarily non-associative and so it is simply a
matter of apples and oranges. On the one hand you have the non-associative
Poisson Algebra (our apples) and on the other hand you have the associative
algebra of observables (our oranges). $Q$ then tries to map apples to
oranges and it seems obvious that there will be inconsistencies in this
mapping. If you run $Q$ only on polynomials that are at most quadratic in $x$
and $p$ then associativity issues never need to come up. However, $Q$ hides
the coordinate invariance of the observable algebra that should result from
the original Poisson algebra.

\subsection{Path Integral Methods}

The path integral, first developed by Feynman, is another equivalent
description of quantum mechanics which is generally covariant. It is based
on an S-matrix which concentrates the focus on how states evolve, i.e., the
propagator $\left\langle x_{f}t_{f}|x_{i}t_{i}\right\rangle $, where $%
x_{i}:=x\left( t_{i}\right) $ and $x_{f}:=x\left( t_{f}\right) $. Starting
with a Lagrangian $L\left( x,\dot{x}\right) $ the propagator can be written: 
\begin{equation*}
\left\langle x_{f}t_{f}|x_{i}t_{i}\right\rangle =N\int Dx~\exp \left[ \frac{1%
}{i\hbar }\int_{t_{i}}^{t_{f}}dt~L\left( x,\dot{x}\right) \right]
\end{equation*}%
Here the sum over all paths is denoted by $\int Dx$ and $N$ is the
normalization constant.

%TCIMACRO{\TeXButton{Witten E. (1988)}{\hyperlink{ref1}{Witten E. (1988)}} }%
%BeginExpansion
\hyperlink{ref1}{Witten E. (1988)}
%EndExpansion
showed that the path integrals on an arbitrary four-dimensional manifold of
a twisted supersymmetric QFT are topological invariants called Donaldson's
polynomial invariants. Thus his model as well as others like it are
diffeomorphism invariant and regarded as topological field theories because
the Hilbert spaces (in a BRST sense) are global topological objects. The
only observables here are those of topological invariants. This established
that the method of path integral quantization is generally covariant and a
major reason of its huge success. The only problem here is that it seems you
are forced to choose your Lagrangian $L$ and the axioms of quantization
should be independent of this choice.

\subsection{Deformation Quantization}

So far we are left with two not-so-appealing options: choose between a
quantization that depends on the coordinates (canonical quantization) or one
that depends on the dynamics (path integral). This brings us to deformation
quantization (DQ).

In 
%TCIMACRO{%
%\TeXButton{Groenewold H. (1946)}{\hyperlink{ref1}{Groenewold H. (19460}} }%
%BeginExpansion
\hyperlink{ref1}{Groenewold H. (19460}
%EndExpansion
(and later in 
%TCIMACRO{\TeXButton{Moyal J. 1949}{\hyperlink{ref1}{Moyal J. 1949}}}%
%BeginExpansion
\hyperlink{ref1}{Moyal J. 1949}%
%EndExpansion
) realized that the Weyl quantization procedure $\mathcal{W}$ along with
Wigner's inverse map $\mathcal{W}^{-1}$ could be used to create an
associative, noncommutative product of the two functions $f$ and $g$ of
phase-space variables defined by $f\ast g:=$ $\mathcal{W}^{-1}\left( 
\mathcal{W}\left( f\right) \mathcal{W}\left( g\right) \right) $ which has
the familiar commutators:%
\begin{equation*}
\left[ x^{\mu },p_{\nu }\right] _{\ast }=i\hbar \delta _{\nu }^{\mu }\text{
\ \ },\text{ \ \ }\left[ x^{\mu },x^{\nu }\right] _{\ast }=0=\left[ p_{\mu
},p_{\nu }\right] _{\ast }
\end{equation*}%
where:%
\begin{equation*}
f\ast g=f\exp \left[ \frac{i\hbar }{2}\left( \frac{\overleftarrow{\partial }%
}{\partial x^{\mu }}\frac{\overrightarrow{\partial }}{\partial p_{\mu }}-%
\frac{\overleftarrow{\partial }}{\partial p_{\mu }}\frac{\overrightarrow{%
\partial }}{\partial x^{\mu }}\right) \right] g
\end{equation*}%
and the arrows denote the direction in which the derivative acts.

In a coordinate independent formulation we have:%
\begin{equation}
f\ast g=f\exp \left[ \overleftrightarrow{P}\right] g  \label{Moyal}
\end{equation}%
\begin{equation*}
\overleftrightarrow{P}:=\overleftarrow{\partial }_{A}\frac{i\hbar }{2}\omega
_{AB}\overrightarrow{\partial }_{B}
\end{equation*}%
where $\overleftrightarrow{P}$ is the Poisson bracket and $\partial _{A}$ is
a (flat) torsion-free phase-space connection ($\partial \otimes \omega =0$).

In summary, what was obtained was another equivalent formulation of the
quantum theory on phase-space, that we call deformation quantization (DQ).
DQ is valid for all phase-space functions and not just ones which are at
most quadratic in position and momenta (see 
%TCIMACRO{%
%\TeXButton{Hancock J. et al 2004}{\hyperlink{ref1}{Hancock J. \textit{et al} 2004}}}%
%BeginExpansion
\hyperlink{ref1}{Hancock J. \textit{et al} 2004}%
%EndExpansion
). Moreover, this is a diffeomorphism covariant quantization which does not
depend on the choice of dynamics (like the Lagrangian in the path integral).

\section{Operator Ordering 
%TCIMACRO{\TeXButton{\\}{\\ }}%
%BeginExpansion
\\ %
%EndExpansion
Ambiguities}

The Weyl quantization map $\mathcal{W}$ on flat space-times corresponds to a
symmetric ordering quantization, e.g.%
\begin{equation*}
\mathcal{W}\left( xp\right) =\frac{1}{2}\left( \hat{x}\hat{p}+\hat{p}\hat{x}%
\right)
\end{equation*}%
\begin{equation*}
\mathcal{W}\left( x^{2}p\right) =\frac{1}{3}\left( \hat{x}^{2}\hat{p}+\hat{x}%
\hat{p}\hat{x}+\hat{p}\hat{x}^{2}\right)
\end{equation*}%
A different ordering choice would correspond to a different quantization
procedure $\mathcal{W}_{\lambda }$ and, in an analogous way, we define a
star-product by (see 
%TCIMACRO{%
%\TeXButton{Hirshfeld A. and Henselder P. 2002}{\hyperlink{ref1}{Hirshfeld A. and Henselder P. 2002}}}%
%BeginExpansion
\hyperlink{ref1}{Hirshfeld A. and Henselder P. 2002}%
%EndExpansion
):%
\begin{equation*}
f\ast _{\lambda }g:=\mathcal{W}_{\lambda }^{-1}\left( \mathcal{W}_{\lambda
}\left( f\right) \mathcal{W}_{\lambda }\left( g\right) \right)
\end{equation*}%
An example of another ordering is standard ordering $\mathcal{W}_{\lambda
}\left( xp\right) :=\hat{x}\hat{p}$ which corresponds to the standard
star-product $\ast _{S}$. In some choice of coordinates $\left( x,p\right) $
it is:%
\begin{equation}
f\ast _{S}g=f\exp \left[ i\hbar \frac{\overleftarrow{\partial }}{\partial
x^{\mu }}\frac{\overrightarrow{\partial }}{\partial p_{\mu }}\right] g
\label{standard}
\end{equation}%
Here we observe that different operator orderings correspond to different
star-products.

Now we have a remarkable theorem:

\textbf{Thm. }\textit{All star-products on a symplectic manifold (a
generalized phase-space) fall into equivalence classes which are
parametrized by a formal series in }$\hbar $\textit{\ with coefficients in
the second de Rham cohomology group }$H_{dR}^{2}\left[ \left[ \hbar \right] %
\right] $\textit{.}

*Note: This theorem is due to the contribution of many people (see 
%TCIMACRO{%
%\TeXButton{Dito G. and Sternheimer D. 2002}{\hyperlink{ref1}{Dito G. and Sternheimer D. 2002}} }%
%BeginExpansion
\hyperlink{ref1}{Dito G. and Sternheimer D. 2002}
%EndExpansion
for a brief history of the classification).

In each equivalence class, whether we describe our system with $\ast _{1}$
or $\ast _{2}$, all physical quantities (like expectation values) will be
identical. For example, the above says that if the magnetic monopole charge
in our space-time is zero then all star-products are equivalent (see 
%TCIMACRO{%
%\TeXButton{Bordemann M. et al 2003}{\hyperlink{ref1}{Bordemann M. \textit{et al} 2003}}}%
%BeginExpansion
\hyperlink{ref1}{Bordemann M. \textit{et al} 2003}%
%EndExpansion
). Additionally, it can be observed in 
%TCIMACRO{%
%\TeXButton{Hirshfeld A. and Henselder P. 2002}{\hyperlink{ref1}{Hirshfeld A. and Henselder P. 2002}} }%
%BeginExpansion
\hyperlink{ref1}{Hirshfeld A. and Henselder P. 2002}
%EndExpansion
several different operator orderings (including the standard on given above $%
\left( \text{%
%TCIMACRO{\TeXButton{\ref{standard}}{\ref{standard}}}%
%BeginExpansion
\ref{standard}%
%EndExpansion
}\right) $) are equivalent to the Groenewold-Moyal star-product $\left( 
\text{%
%TCIMACRO{\TeXButton{\ref{Moyal}}{\ref{Moyal}}}%
%BeginExpansion
\ref{Moyal}%
%EndExpansion
}\right) $. In other words there are examples of operator orderings that do
not effect the physics. Thus, the task of understanding how an arbitrary
operator ordering affects the physics is now reduced to analyzing these
equivalence classes.

\section{The Klein-Gordon 
%TCIMACRO{\TeXButton{\\}{\\ }}%
%BeginExpansion
\\ %
%EndExpansion
Equation and the Fedosov Star-Product}

In order to gain a basic feel for DQ, we will recast the well known equation
Klein-Gordon (KG) equation into DQ. In this section we will sometimes
implicitly use $\mathcal{W}$ and $\mathcal{W}^{-1}$ to go from Hilbert
spaces to phase-space (and back). For more details of the arguments below,
see 
%TCIMACRO{%
%\TeXButton{Tillman P. and Sparling G. (2006a, 2006b)}{\hyperlink{ref1}{Tillman P. and Sparling G. (2006a, 2006b)}}}%
%BeginExpansion
\hyperlink{ref1}{Tillman P. and Sparling G. (2006a, 2006b)}%
%EndExpansion
.

The KG equation is obtained by promoting the classical Minkowskian
relativistic invariant $p_{\mu }p^{\mu }-m^{2}$ to a Hilbert space operator.

States of definite mass $\left\vert \phi _{m}\right\rangle $ are then
solutions to the eigenvalue equation:%
\begin{equation*}
(\hat{p}_{\mu }\hat{p}^{\mu }-m^{2})\left\vert \phi _{m}\right\rangle
=0~~~,~~~\left\langle \phi _{m}|\phi _{m}\right\rangle =1
\end{equation*}%
This is the KG equation because in $x$-space we have:%
\begin{equation*}
(\partial _{\mu }\partial ^{\mu }+m^{2}/\hbar ^{2})\phi _{m}\left( x\right)
=0
\end{equation*}%
To reformulate states as quantities in phase-space (i.e. in DQ) we use
Wigner's inverse map $\mathcal{W}^{-1}$:%
\begin{equation*}
\rho _{m}:=\mathcal{W}^{-1}\left( \left\vert \phi _{m}\rangle \langle \phi
_{m}\right\vert \right)
\end{equation*}%
The functions $\rho _{m}$ are known as Wigner functions.

The KG equation on Minkowski space in DQ can be now written as:%
\begin{equation*}
H\ast \rho _{m}=\rho _{m}\ast H=m^{2}\rho _{m}
\end{equation*}%
\begin{equation*}
H=p_{\mu }\ast p^{\mu }
\end{equation*}%
\begin{equation*}
Tr_{\ast }\left( \rho _{m}\right) =1~~~,~~~\bar{\rho}_{m}=\rho _{m}
\end{equation*}%
where $\ast $ is the Groenewold-Moyal star-product and $Tr_{\ast }$ is the
trace over all degrees of freedom.

In an analogous derivation (and by adding an arbitrary Ricci term) we can
formulate the KG equation on an arbitrary space-time:%
\begin{equation}
H\ast \rho _{m}=\rho _{m}\ast H=m^{2}\rho _{m}  \label{KG}
\end{equation}%
\begin{equation}
H=p_{\mu }\ast p^{\mu }+\xi R\left( x\right)  \label{H}
\end{equation}%
\begin{equation*}
Tr_{\ast }\left( \rho _{m}\right) =1~~~,~~~\bar{\rho}_{m}=\rho _{m}
\end{equation*}%
where $\ast $ is now the Fedosov\ star-product (a generalization of the
Groenewold-Moyal star-product), $g_{\mu \nu }\left( x\right) $ is a
configuration space metric, $R\left( x\right) $ is the Ricci scalar
associated to this metric, $p^{\mu }:=g^{\mu \nu }p_{\nu }$, and $\xi \in 
%TCIMACRO{\U{2102} }%
%BeginExpansion
\mathbb{C}
%EndExpansion
$ is an arbitrary constant.\pagebreak

The properties of the Fedosov star are (see 
%TCIMACRO{\TeXButton{Fedosov B. 1996}{\hyperlink{ref1}{Fedosov B. 1996}}}%
%BeginExpansion
\hyperlink{ref1}{Fedosov B. 1996}%
%EndExpansion
, 

\noindent 
%TCIMACRO{%
%\TeXButton{Tillman P. and Sparling G. (2006a, 2006b)}{\hyperlink{ref1}{Tillman P. and Sparling G. (2006a, 2006b)}}}%
%BeginExpansion
\hyperlink{ref1}{Tillman P. and Sparling G. (2006a, 2006b)}%
%EndExpansion
):

\begin{enumerate}
\item It is diffeomorphism covariant.

\item It can be constructed on all symplectic manifolds (including all
phase-spaces) perturbatively in powers of $\hbar $.

\item It assumes no dynamics (e.g. Hamiltonian or Lagrangian), symmetries,
or even a metric.

\item The limit $\hbar \rightarrow 0$ yields classical mechanics.

\item It is equivalent to an operator formalism by a Weyl-like quantization
map $\sigma ^{-1}$.
\end{enumerate}

The Fedosov\ star-product is given by an iterative construction, and, with
convergence issues aside, all star products on any symplectic manifold\ are
formally equivalent to a Fedosov star (see 
%TCIMACRO{%
%\TeXButton{Dito G. and Sternheimer D. 2002}{\hyperlink{ref1}{Dito G. and Sternheimer D. 2002}}}%
%BeginExpansion
\hyperlink{ref1}{Dito G. and Sternheimer D. 2002}%
%EndExpansion
). We add that the role played $\mathcal{W}$\ (and $\mathcal{W}^{-1}$) is
the flat section in Weyl bundle (called $\sigma ^{-1}$ in 
%TCIMACRO{\TeXButton{Fedosov B. 1996}{\hyperlink{ref1}{Fedosov B. 1996}}}%
%BeginExpansion
\hyperlink{ref1}{Fedosov B. 1996}%
%EndExpansion
) over the symplectic manifold.

\subsection{The dS and AdS Space-Times}

We constructed the Fedosov star-product for the phase-space a class of
constant curvature manifolds in 
%TCIMACRO{%
%\TeXButton{Tillman P. and Sparling G. (2006a)}{\hyperlink{ref1}{Tillman P. and Sparling G. (2006a)}}}%
%BeginExpansion
\hyperlink{ref1}{Tillman P. and Sparling G. (2006a)}%
%EndExpansion
. The following is a summary of these results for the cases of the dS and
AdS space-times.

One of the goals of these results is to obtain a nonperturbative
construction of the Fedosov star-product for the dS/AdS space-times. Another
is to verify that the algebra of observables and the KG equation reproduced
previous results of 
%TCIMACRO{%
%\TeXButton{Fr\o nsdal C. (1965, 1973, 1975a, 1975b)}{\hyperlink{ref1}{Fr\o nsdal C. (1965, 1973, 1975a, 1975b)}}}%
%BeginExpansion
\hyperlink{ref1}{Fr\o nsdal C. (1965, 1973, 1975a, 1975b)}%
%EndExpansion
.

We first embed dS/AdS in a flat five dimensional space given by the
embedding formulas:%
\begin{equation*}
\eta _{\mu \nu }x^{\mu }x^{\nu }=1/C\text{ \ \ and \ \ }x^{\mu }p_{\mu }=A
\end{equation*}

where $C$ and $A$ are some real arbitrary constants, and $\eta $ is the
embedding flat metric. For dS $\eta =diag\left( 1,-1,-1,-1,-1\right) $, $C<0$
and AdS $\eta =diag\left( 1,1,-1,-1,-1\right) $, $C>0$.

For brevity we omit the technical details of the calculations and simply
give the results. We obtain the exact results for the Fedosov
star-commutators:%
\begin{equation}
\left[ x^{\mu },x^{\nu }\right] _{\ast }=0~~\ ~~[x_{\mu },M_{\nu \rho
}]_{\ast }=i\hbar x_{[\nu }\eta _{\rho ]\mu }
\end{equation}%
\begin{equation*}
\lbrack M_{\mu \nu },M_{\rho \sigma }]_{\ast }=i\hbar (M_{\rho \lbrack \mu
}\eta _{\nu ]\sigma }-M_{\sigma \lbrack \mu }\eta _{\nu ]\rho })
\end{equation*}%
indices run from $0$ to $4$, $M_{\mu \nu }=x_{[\mu }\ast p_{\nu ]}$, $x_{\mu
}=\eta _{\mu \nu }x^{\nu }$.

The conditions of the embedding $x^{\mu }x_{\mu },~x^{\mu }p_{\mu }$ become
the Casimir invariants of the algebra in group theoretic language.

We now summarize our two key observations:

\begin{enumerate}
\item $M$'s generate $\mathbb{SO}\left( 1,4\right) $ and $\mathbb{SO}\left(
2,3\right) $ for dS and AdS respectively.

\item $M$'s and $x$'s generate $\mathbb{SO}\left( 2,4\right) $ for \textit{%
both} dS and AdS.
\end{enumerate}

By calculating $R=-16C$ and $p_{\mu }\ast p^{\mu }$ in terms of $M$ and $x$
the Hamiltonian $\left( \text{%
%TCIMACRO{\TeXButton{\ref{H}}{\ref{H}}}%
%BeginExpansion
\ref{H}%
%EndExpansion
}\right) $ is:%
\begin{equation}
H=2CM_{\mu \nu }\ast M^{\mu \nu }+\left( A-4i\hbar \right) AC-16\xi C
\end{equation}%
where $M_{\mu \nu }\ast M^{\mu \nu }$ is a Casimir invariant of the subgroup 
$\mathbb{SO}\left( 1,4\right) $ or $\mathbb{SO}\left( 2,3\right) $ for dS or
AdS respectively.

In the more familiar form of Hilbert space language the KG equation $\left( 
\text{%
%TCIMACRO{\TeXButton{\ref{KG}}{\ref{KG}}}%
%BeginExpansion
\ref{KG}%
%EndExpansion
}\right) $ takes the form:%
\begin{equation}
(2C\hat{M}_{\mu \nu }\hat{M}^{\mu \nu }+\chi C)\left\vert \phi
_{m}\right\rangle =m^{2}\left\vert \phi _{m}\right\rangle  \label{KG1}
\end{equation}%
where $\left\langle \phi _{m}|\phi _{m}\right\rangle =1$, $%
%TCIMACRO{\U{2102} }%
%BeginExpansion
\mathbb{C}
%EndExpansion
\ni \chi =\left( A-4i\hbar \right) A-16\xi $ is an arbitrary constant, and
we regard all groups to be in a standard irreducible representation on the
set of linear Hilbert space operators.

These subgroups are the symmetry groups of the manifolds for dS or AdS
respectively. Again, $\hat{M}_{\mu \nu }\hat{M}^{\mu \nu }$ is a Casimir
invariant of the subgroup $\mathbb{SO}\left( 1,4\right) $ or $\mathbb{SO}%
\left( 2,3\right) $ for dS or AdS respectively. Therefore, the above KG
equation $\left( \text{%
%TCIMACRO{\TeXButton{\ref{KG1}}{\ref{KG1}}}%
%BeginExpansion
\ref{KG1}%
%EndExpansion
}\right) $ states that the eigenstates of mass $\left\vert \phi
_{m}\right\rangle $ label the different representations of $\mathbb{SO}%
\left( 1,4\right) $ and $\mathbb{SO}\left( 2,3\right) $ for dS and AdS
respectively sitting inside the full group of observables $\mathbb{SO}\left(
2,4\right) $ which is confirmed by 
%TCIMACRO{%
%\TeXButton{Fr\o nsdal C. (1965, 1973)}{\hyperlink{ref1}{Fr\o nsdal C. (1965, 1973)}} }%
%BeginExpansion
\hyperlink{ref1}{Fr\o nsdal C. (1965, 1973)}
%EndExpansion
as well as others.

\section{Conclusions}

As we saw, the main advantage over both canonical quantization and path
integral methods is that DQ\ is \textit{both} diffeomorphism covariant and
independent of the dynamics (e.g. Lagrangian). Also, the operator ordering
ambiguity is reduced to the task of analyzing the equivalence classes of
star-products in the context of DQ by knowing two facts: different operator
ordering corresponds to different star-products and the classification of
all star-products on a symplectic manifold. Additionally we mentioned an
example of two equivalent orderings: standard and symmetric (Weyl); there
are more examples of orderings that yield equivalent star-products in 
%TCIMACRO{%
%\TeXButton{Hirshfeld A. and Henselder P. (2002)}{\hyperlink{ref1}{Hirshfeld A. and Henselder P. (2002)}}}%
%BeginExpansion
\hyperlink{ref1}{Hirshfeld A. and Henselder P. (2002)}%
%EndExpansion
.

It is shown how to conceptually move from a Hilbert space formalism to DQ
and back using implicitly the map $\mathcal{W}$\ and its generalization $%
\sigma ^{-1}$. This helped us reformulate quantities and equations, like the
KG equation, from DQ into Hilbert spaces (and back). For the specific cases
of dS and AdS space-times the Fedosov star-product was calculated and the
results obtained were the expected ones (see 
%TCIMACRO{%
%\TeXButton{Fr\o nsdal C. 1965, 1973, 1975a, 1975b}{\hyperlink{ref1}{Fr\o nsdal C. 1965, 1973, 1975a, 1975b}}}%
%BeginExpansion
\hyperlink{ref1}{Fr\o nsdal C. 1965, 1973, 1975a, 1975b}%
%EndExpansion
). However, the fundamental issue of convergence of all formal series in DQ
still remains unknown.

\section*{Acknowledgements}

I would like to thank my advisor George Sparling for very helpful
discussions.

\section*{References}

\noindent 
%TCIMACRO{\TeXButton{\hypertarget{ref1}{}}{\hypertarget{ref1}{}}}%
%BeginExpansion
\hypertarget{ref1}{}%
%EndExpansion
Bordemann M. \textit{et al} 2003 \textit{J. Funct. Anal. }\textbf{199} (1),
1.

\noindent Dito G. 2002 \textit{Proc. Int. Conf. of 68}$^{\text{\textit{\`{e}%
me}}}$\textit{\ Rencontre entre Physiciens Th\'{e}oriciens et Math\'{e}%
maticiens on Deformation Quantization I (Strasbourg)}, (Berlin: de Gruyter)
p 55 \textit{Preprint} 
%TCIMACRO{%
%\TeXButton{math.QA/0202271}{\href{http://xxx.lanl.gov/abs/math.QA/0202271}{math.QA/0202271}}}%
%BeginExpansion
\href{http://xxx.lanl.gov/abs/math.QA/0202271}{math.QA/0202271}%
%EndExpansion
.

\noindent Dito G. and Sternheimer D. 2002 \textit{Proc. Int. Conf. of 68}$^{%
\text{\textit{\`{e}me}}}$\textit{\ Rencontre entre Physiciens Th\'{e}%
oriciens et Math\'{e}maticiens on Deformation Quantization I (Strasbourg)},
(Berlin: de Gruyter) p 9 \textit{Preprint} 
%TCIMACRO{%
%\TeXButton{math.QA/0201168}{\href{http://xxx.lanl.gov/abs/math.QA/0201168}{math.QA/0201168}}}%
%BeginExpansion
\href{http://xxx.lanl.gov/abs/math.QA/0201168}{math.QA/0201168}%
%EndExpansion
.

\noindent Fedosov\ B. 1996\textit{\ Deformation Quantization and Index Theory%
} (Berlin: Akademie).

\noindent Fr\o nsdal C. 1965 \textit{Rev. Mod. Phys.} \textbf{37}, 221.

\noindent Fr\o nsdal C. 1973 \textit{Phys. Rev.} D \textbf{10}, 2, 589.

\noindent Fr\o nsdal C. 1975a \textit{Phys. Rev.} D \textbf{12}, 12, 3810.

\noindent Fr\o nsdal C. 1975b \textit{Phys. Rev.} D \textbf{12}, 12, 3819.

\noindent Giulini D. 2003 \textit{Quantum Gravity: From Theory to
Experimental Search}, (Bad Honnef, Germany: Springer) p 17-40.

\noindent Gozzi E. and Reuter M. 1994 \textit{Int. J.Mod.Phys.} \textbf{A9},
32, 5801.

\noindent Groenewold H. J. 1946 \textit{Physica} \textbf{12}, 405.

\noindent Hirshfeld A. and Henselder P. 2002 \textit{Am. J. Phys.} \textbf{70%
} (5), 537.

\noindent Hancock\ J. \textit{et al} 2004 \textit{Eur. J. Phys.} \textbf{25,}
525.

\noindent Moyal J.E. 1949 \textit{Proc. Cambridge Phil. Soc.} \textbf{45},
99.

\noindent Tillman P. and Sparling G. 2006a Fedosov Observables on Constant
Curvature Manifolds and the Klein-Gordon Equation, \textit{Preprint} 
%TCIMACRO{%
%\TeXButton{gr-qc/0603017}{\href{http://xxx.lanl.gov/abs/gr-qc/0603017}{gr-qc/0603017}}}%
%BeginExpansion
\href{http://xxx.lanl.gov/abs/gr-qc/0603017}{gr-qc/0603017}%
%EndExpansion
.

\noindent Tillman P. and Sparling G. 2006b \textit{J. Math. Phys.} \textbf{47%
}, 052102.

\noindent Witten E. 1988 \textit{Commun. Math. Phys.} \textbf{117}, 353.

\noindent Woodhouse\ N. 1980\textit{\ Geometric Quantization} (New York:
Oxford University Press).

%TCIMACRO{\TeXButton{\end{multicols}}{\end{multicols}}}%
%BeginExpansion
\end{multicols}%
%EndExpansion

\end{document}